\documentclass[onecolumn,showpacs,preprintnumbers,amsmath,amssymb]{revtex4}
\usepackage{graphicx}% Include figure files
\usepackage{dcolumn}% Align table columns on decimal point
\usepackage{bm}% bold math

\begin{document}

%\preprint{APS/123-QED}

\title{Electromagnetic dissociation of relativistic $^8$B nuclei in nuclear track emulsion}% Force line breaks with \\
\author{R.~Stanoeva}
     \email{stanoeva@lhe.jinr.ru}
     \homepage{http://becquerel.jinr.ru}
   \affiliation{Joint Insitute for Nuclear Research, Dubna, Russia}
\author{D.~A.~Artemenkov}
   \affiliation{Joint Insitute for Nuclear Research, Dubna, Russia}
\author{V.~Bradnova}
   \affiliation{Joint Insitute for Nuclear Research, Dubna, Russia}
\author{S.~Vok\'al}
   \affiliation{P. J. \u Saf\u arik University, Ko\u sice, Slovak Republic}
\author{L.~A.~Goncharova}
   \affiliation{Lebedev Institute of Physics, Russian Academy of Sciences, Moscow, Russia}
\author{P.~I.~Zarubin}
   \affiliation{Joint Insitute for Nuclear Research, Dubna, Russia}
\author{I.~G.~Zarubina}
   \affiliation{Joint Insitute for Nuclear Research, Dubna, Russia}
\author{N.~A.~Kachalova}
   \affiliation{Joint Insitute for Nuclear Research, Dubna, Russia}
\author{A.~D.~Kovalenko}
   \affiliation{Joint Insitute for Nuclear Research, Dubna, Russia}
\author{D.~O.~Krivenkov}
   \affiliation{Joint Insitute for Nuclear Research, Dubna, Russia}
\author{A.~I.~Malakhov}
   \affiliation{Joint Insitute for Nuclear Research, Dubna, Russia}
\author{G.~I.~Orlova}
   \affiliation{Lebedev Institute of Physics, Russian Academy of Sciences, Moscow, Russia}
\author{N.~G.~Peresadko}
   \affiliation{Lebedev Institute of Physics, Russian Academy of Sciences, Moscow, Russia}
\author{N.~G.~Polukhina}
   \affiliation{Lebedev Institute of Physics, Russian Academy of Sciences, Moscow, Russia}
\author{P.~A.~Rukoyatkin}
   \affiliation{Joint Insitute for Nuclear Research, Dubna, Russia}
\author{V.~V.~Rusakova}
   \affiliation{Joint Insitute for Nuclear Research, Dubna, Russia}
\author{M.~Haiduc}
   \affiliation{Institute of Space Sciences, Magurele, Romania}
\author{S.~P.~Kharlamov}
   \affiliation{Lebedev Institute of Physics, Russian Academy of Sciences, Moscow, Russia}
\author{M.~M.~Chernyavsky}
   \affiliation{Lebedev Institute of Physics, Russian Academy of Sciences, Moscow, Russia}
\author{T.~V.~Shchedrina}
   \affiliation{Joint Insitute for Nuclear Research, Dubna, Russia}

\date{\today}% It is always \today, today,

\begin{abstract}
\indent Experimental data on fragmentation channels in peripheral interactions of $^8$B nuclei in nuclear track emulsions are presented. A detailed analysis made it possible to justify selections of events of
the electromagnetic-dissociation process $^8$B $\rightarrow^7$Be + \emph{p} and to estimate its cross section. Events of $^{10}$C peripheral dissociation that were observed in the same exposure are described.\par
\end{abstract}


\maketitle

\section{\label{sec:level1} INTRODUCTION}

\par\bigskip

As is well known, the advent of beams of radioactive nuclei opens qualitatively new possibilities for studying the structural features of such nuclei and their excited states (for a recent overview, see \cite{Aumann05}). Of particular interest are peripheral interactions at an energy of about 1 GeV per nucleon since they are optimal in measurement conditions and interpretation. In the present article, we report on an investigation of the fragmentation of $^8$B nuclei in a nuclear track emulsion at a projectile energy of 1.2 GeV per nucleon. This investigation relies on the potential of the nuclotron of the Joint Institute for Nuclear Research (JINR, Dubna) for creating beams of relativistic light nuclei including radioactive nuclei. It is of interest to study the properties of the $^7$Be + \emph{p} system, which is close to the ground state of the $^8$B nucleus, as well as its strongly excited states as the three-center system of $^{1,2}$H, $^{3,4}$He, and $^6$Li nucleon clusters. Possibly, investigation of three-cluster features of the structure of the $^8$B nucleus will furnish sufficient grounds to supplement nucleosynthesis in fast processes with the three-body fusion process $^3$He + $^2$H + $^3$He. The above cluster configurations must manifest themselves in the exclusive observation of dissociation channels near the corresponding thresholds.

By and large, peripheral reactions are intricate because of the interplay of various mechanisms, including electromagnetic and nuclear diffractive interactions, as well as nucleon-stripping reactions. In the most peripheral collisions, participant nuclei interact with each other via time-dependent electromagnetic fields (for a theoretical review on the subject, see \cite{Bertulani88,Baur96,nuclth/9710060,Baur03}), and this makes it possible to study the interaction of nuclei with quasireal target photons or even with their coherent groups (multiphoton processes). The intensity of equivalent-photon spectra increases with increasing beam energy. The Coulomb interaction may excite states of ever higher energy, and this leads to the opening of new nuclear-dissociation channels in addition to excitations that do not break the binding of nucleons. Radioactive nuclei can only be studied in their secondary beams.

\begin{figure}
\includegraphics[width=160mm]{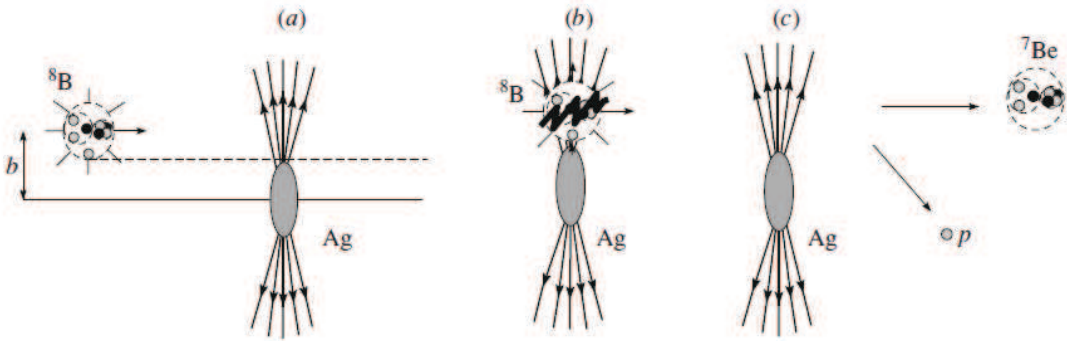}
\caption{\label{fig:1} Scheme of the electromagnetic dissociation of a relativistic $^8$B nucleus in the field of a silver nucleus: (\emph{a}) approach of the nuclei at an impact parameter \emph{b}; (\emph{b}) quasireal-photon absorption by the $^8$B nucleus; and (\emph{c}) dissociation into two fragments, \emph{p} and $^7$Be.}
\end{figure}

Despite interest in the full pattern of relativistic fragmentation, experimental advances in these realms are hindered by a number of problems. In any detector type, an increase in the degree of dissociation of a relativistic nucleus leads to a drastic reduction of the fragment-ionization signal in proportion to the square of the fragment charge. This circumstance
complicates the detection of relativistic fragments up to helium and a hydrogen isotopes. Owing to an extremely small binding energy of the outer proton, the $^8$B nucleus is the most sensitive indicator of electromagnetic
interaction with heavy nuclei (Fig. \ref{fig:1}). However, the number of studies in which a proton was detected in addition to a $^7$Be nucleus is quite limited. For example, $^8$B photodissociation through the $^7$Be + \emph{p} channel was studied in a GSI counter experiment at a beam energy of up to 250 MeV per nucleon \cite{Iwasa99}. More complicated channels escape detection in such experiments because of technical limitations.

The method of nuclear track emulsions is the most appropriate for studying the relativistic fragmentation of neutron-deficient light nuclei, because this method ensures the most detailed observation of the interaction pattern (examples illustrating the application of the method can be found in \cite{web,Adamovich99,Andreeva05}). The traditional task of the nuclear-track-emulsion method is to outline the nuclear-interaction pattern on the basis of a limited statistical data sample in order to plan better future complicated experiments featuring various detectors. Limitations on the statistics subjected to analysis are compensated to some extent by the impossibility of completely observing the composition of fragments within other methods.

The presence of rather heavy nuclei of silver and bromine and a group of light nuclei of carbon, nitrogen, oxygen, and hydrogen in similar concentrations in nuclear track emulsions is of use in comparing peripheral interactions of different types. Under identical conditions, one can observe the breakup of a nucleus both in the electromagnetic field of a heavy target nucleus and in collisions with target protons. Events of the fragmentation of track-emulsion nuclei contain strongly ionizing target fragments, including
alpha particles, protons of energy below 26 MeV, and light recoil nuclei (\emph{b} particles, whose multiplicity is denoted by \emph{n}$_b$), as well as nonrelativistic protons of energy in excess of 26 MeV (\emph{g} particles, whose multiplicity is denoted by \emph{n}$_g$). Relevant reactions are also characterized by the multiplicity \emph{n}$_s$ of product mesons (\emph{s} particles). Using these parameters, one can draw preliminary conclusions on the character of the interaction.

The $^8$B nucleus is first studied by the method of nuclear track emulsion. This makes it possible to deduce information both about the dissociation
channel $^8$B $\rightarrow^7$Be + \emph{p} and about dissociation into extremely light nuclei He and H. Microphotographs of such events were presented in \cite{Artemenkov08}. Of particular interest are events that do not contain target-nucleus fragments or charged mesons (\emph{n}$_b$ = 0, \emph{n}$_g$ = 0, \emph{n}$_s$ = 0)$-$these are so-called white stars. They form an observational basis in searches for electromagnetic-dissociation events. The objective of present study is to seek white stars that are formed at the minimum momentum transfer and which could be associated with the electromagnetic dissociation of $^8$B nuclei.

A theoretical analysis of the contributions of various mechanisms to the cross section for the reaction $^8$B $\rightarrow^7$Be + \emph{p} of proton separation from the $^8$B nucleus was performed in \cite{Esbensen2000}, where predictions were made for the behavior of the cross section as a function of the energy of the $^8$B nucleus in the region extending up to about 2 GeV per nucleon and for the branching fractions of Coulomb dissociation, nuclear dissociation, and the stripping reaction for the case of a Pb target. All of these mechanisms lead to the maximum cross section at an impact parameter of
\emph{b} $\approx$ 10 fm $-$ that is, in the case of closest approach of participant nuclei without overlap of the densities. The contribution of the stripping reaction to the cross section is approximately five times as great as the
contributions of Coulomb and nuclear diffraction, but it decreases sharply with increasing \emph{b}. The nuclear diffraction contribution decreases at the same rate with increasing \emph{b}. At impact-parameter values in the range \emph{b} = 12-15 fm, Coulomb dissociation becomes dominant. The integrated cross sections at 1.2 GeV per nucleon are about 210 (Coulomb interaction), 190 (stripping) and 50 mb (nuclear diffraction). Stripping reactions are observed with a high efficiency in trackemulsion experiments. In the interaction with a target, the protons of the $^8$B nucleus generate secondary fragments and mesons or undergo a strong deflection, and one can use this to exclude the stripping contribution.

We will use our data to estimate the cross sections for electromagnetic and diffractive dissociation on track-emulsion nuclei. In the case of dissociation on Ag and Br nuclei, the Coulomb cross section decreases to 70 and 40 mb, respectively. Under the assumption of the \emph{A}$_t^{1/3}$ dependence, the nuclear diffraction cross section decreases only to 40 and 36 mb. These mechanisms of interaction on a silver nucleus become competing. However, the use of silver nuclei as an electromagnetic target is motivated by the properties of the track-emulsion procedure. The separation of the nuclear-diffraction contribution becomes quite important. In the case of Coulomb dissociation, breakup occurs on a nearly massless photon rather than on a massive nucleus, and the jet of fragments should receive a minimum recoil. An a extremely small value of the total-transverse momentum transfer to the fragmenting system is an important condition for separating electromagnetic interactions.

\section{\label{sec:level2} CHARGE COMPOSITION OF RELATIVISTIC FRAGMENTS}

\begin{figure}
\includegraphics[width=120mm]{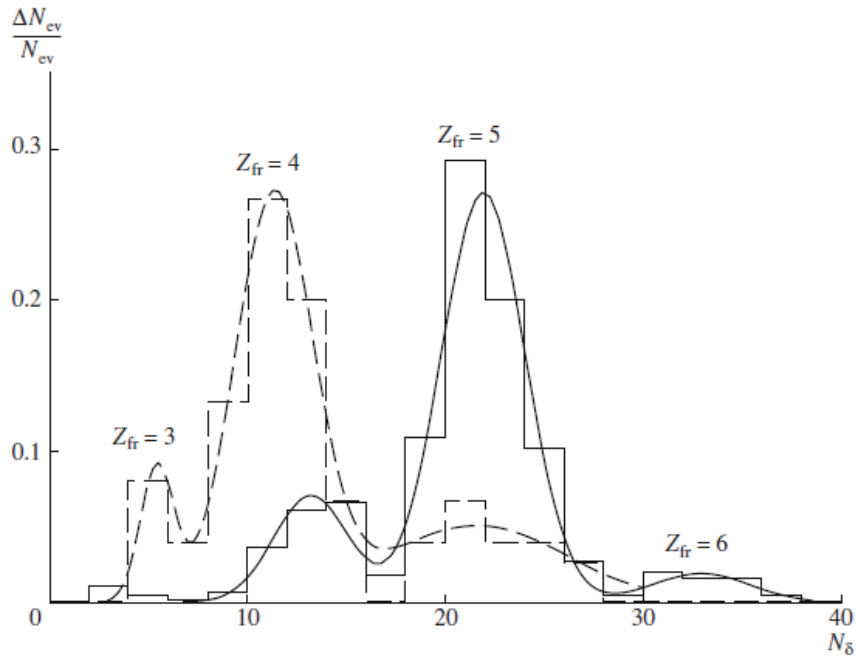}
\caption{\label{fig:2} Distribution of the number of $\delta$-electrons over 1 mm of length of tracks of beam particles that induced interactions under study (440 events; solid-line histogram) and relativistic fragments of charge number \emph{Z}$_{fr}$ in excess of two (75 events; dashed-line histogram). The displayed curves represent approximations by the sum of Gaussian functions.}
\end{figure}

The present study is a continuation of our analysis reported in \cite{Stanoeva07}. New results were obtained in determining the charges of beam nuclei and their relativistic fragments. Details of the exposure of the track emulsion to a secondary beam of relativistic nuclei $^8$B having a momentum of \emph{P$_0$} = 2.0 GeV/\emph{c} per nucleon and a primary analysis of the charge topology of relativistic fragments in the dissociation of $^8$B nuclei are discussed in \cite{Stanoeva07,Rukoyatkin06}. In \cite{Stanoeva07}, a complicated charge composition of the beam prevented an unambiguous identification of the charges of primary nuclei (Z$_{pr}$) by the sum of fragment charges ($\Sigma$\emph{Z}$_{fr}$) alone. For this reason, measurements of the charges \emph{Z}$_{pr}$ by the method of counting $\delta$ electrons (\emph{N}$_\delta$) were performed on all tracks of beam nuclei that induced peripheral interactions characterized by $\Sigma$\emph{Z}$_{fr} > $ 2. The results stemming from the determination of charges on beam-particle tracks and illustrating the accuracy of this method are given in Fig. \ref{fig:2}. The mean value $\langle$\emph{N}$_\delta\rangle$ depends linearly on \emph{Z}$_{pr}^{2}$. The distribution shows quite a distinct separation of nuclei in charge, and this makes it possible to find that Li, Be, B, and C nuclei in the beam are in the ratio 0.01/0.19/0.76/0.04. These results agree with the data from the scintillation beam monitor. The same method was used to obtain the distribution of charges (\emph{Z}$_{fr}$) of 75 secondary fragments (Fig. \ref{fig:2}). The expected change in the distribution is observed.

In the track emulsion irradiated with a beam of the above composition, we found 1070 interactions of \emph{Z}$_{pr}\geq$ 3 nuclei over the total track length of \emph{L} = 144 m. The mean free path before interaction was $\lambda$ = 13.5$\pm$0.4 cm, which corresponds to the expectation based on data for the closest cluster nuclei. In this sample, we selected 434 peripheral-fragmentation stars (\emph{N}$_{pf}$), in which the total charge of relativistic fragments in the fragmentation cone of critical angle 8$^{\circ}$ satisfies the condition $\Sigma$\emph{Z}$_{fr} > $ 2. The statistical sample of \emph{N}$_{pf}$ was broken down into two groups: that of events accompanied by target-nucleus fragments or by product mesons (or by both of them), \emph{N}$_{tf}$, and that of white stars, \emph{N}$_{ws}$. A comparison of these two groups gives a clearer idea of special features of white-star production.

Among peripheral-fragmentation events, 320 interactions have the primary-nucleus charge \emph{Z}$_{pr}$ = 5 and the total charge of particles emitted within the 8$^{\circ}$ cone, $\Sigma$\emph{Z}$_{fr}$, in excess of two. For these events, the distribution over configurations of \emph{N}$_z$ fragments with charges \emph{Z}$_{fr}$ is given in Table \ref{tab:1}. A dominant contribution comes from $\Sigma$\emph{Z}$_{fr}$ = 5 events. The main difference in the statistics of \emph{N}$_{tf}$ and \emph{N}$_{ws}$ events for this group manifests itself in the $\Sigma$\emph{Z}$_{fr}$ = 4+1 two-particle channel, which is interpreted as $^8$B $\rightarrow^7$Be + \emph{p}. Its fraction increases sharply upon selecting white stars, from 13\% for \emph{N}$_{tf}$ to 48\% for \emph{N}$_{ws}$, and this may be due to
$^8$B dissociation through the channel having the lowest threshold.

\begin{table}
\caption{\label{tab:1} Distribution of peripheral interactions (\emph{N}$_{pf}$) over the charge topology of relativistic fragments ($\Sigma$\emph{Z}$_{fr}$) for \emph{Z}$_{pr}$ = 5 primary nuclei (\emph{N}$_{ws}$ is the number of white stars, and \emph{N}$_{tf}$ is the number of events containing target-nucleus fragments)}

\begin{tabular}{l|ccccc|cc}
\hline\noalign{\smallskip}
~~$\Sigma$\emph{Z}$_{fr}$~~&  &  & ~~\emph{N}$_z$~~ & & & ~~\emph{N}$_{ws}$~~ & ~~\emph{N}$_{tf}$~~\\
 & ~~5~~ & ~~4~~ & ~~3~~ & ~~2~~ & ~~1~~ & \\
\noalign{\smallskip}\hline\noalign{\smallskip}
~~7~~ & ~~-~~ & ~~-~~ & ~~-~~ & ~~3~~ & ~~1~~ & ~~-~~ & ~~1~~ \\
~~7~~ & ~~-~~ & ~~-~~ & ~~-~~ & ~~2~~ & ~~3~~ & ~~-~~ & ~~1~~ \\
~~6~~ & ~~-~~ & ~~-~~ & ~~-~~ & ~~2~~ & ~~2~~ & ~~1~~ & ~~12~~\\
~~6~~ & ~~-~~ & ~~-~~ & ~~-~~ & ~~1~~ & ~~4~~ & ~~4~~ & ~~7~~ \\
~~6~~ & ~~-~~ & ~~-~~ & ~~-~~ & ~~-~~ & ~~6~~ & ~~1~~ & ~~2~~ \\
~~6~~ & ~~-~~ & ~~1~~ & ~~-~~ & ~~1~~ & ~~-~~ & ~~-~~ & ~~1~~ \\
~~6~~ & ~~-~~ & ~~1~~ & ~~-~~ & ~~-~~ & ~~2~~ & ~~-~~ & ~~4~~ \\
~~6~~ & ~~1~~ & ~~-~~ & ~~-~~ & ~~-~~ & ~~1~~ & ~~1~~ & ~~2~~ \\
~~5~~ & ~~-~~ & ~~-~~ & ~~-~~ & ~~1~~ & ~~3~~ & ~~12~~& ~~42~~\\
~~5~~ & ~~-~~ & ~~-~~ & ~~-~~ & ~~2~~ & ~~1~~ & ~~14~~& ~~44~~\\
~~5~~ & ~~-~~ & ~~-~~ & ~~1~~ & ~~-~~ & ~~2~~ & ~~-~~ & ~~5~~ \\
~~5~~ & ~~-~~ & ~~-~~ & ~~1~~ & ~~1~~ & ~~-~~ & ~~-~~ & ~~2~~ \\
~~5~~ & ~~-~~ & ~~1~~ & ~~-~~ & ~~-~~ & ~~1~~ & ~~25~~& ~~16~~\\
~~5~~ & ~~1~~ & ~~-~~ & ~~-~~ & ~~-~~ & ~~-~~ & ~~1~~ & ~~13~~\\
~~5~~ & ~~-~~ & ~~-~~ & ~~-~~ & ~~-~~ & ~~5~~ & ~~-~~ & ~~2~~ \\
~~4~~ & ~~-~~ & ~~-~~ & ~~-~~ & ~~-~~ & ~~4~~ & ~~-~~ & ~~17~~\\
~~4~~ & ~~-~~ & ~~-~~ & ~~-~~ & ~~2~~ & ~~-~~ & ~~-~~ & ~~16~~\\
~~4~~ & ~~-~~ & ~~-~~ & ~~-~~ & ~~1~~ & ~~2~~ & ~~1~~ & ~~45~~\\
~~4~~ & ~~-~~ & ~~1~~ & ~~-~~ & ~~-~~ & ~~-~~ & ~~-~~ & ~~8~~ \\
~~4~~ & ~~-~~ & ~~-~~ & ~~1~~ & ~~-~~ & ~~1~~ & ~~-~~ & ~~1~~ \\
~~3~~ & ~~-~~ & ~~-~~ & ~~-~~ & ~~1~~ & ~~1~~ & ~~-~~ & ~~11~~\\
~~3~~ & ~~-~~ & ~~-~~ & ~~-~~ & ~~-~~ & ~~3~~ & ~~-~~ & ~~5~~ \\
\noalign{\smallskip}\hline

\end{tabular}
%\hspace*{10cm}  % with the correct table height
\end{table}

Among $\Sigma$\emph{Z}$_{fr}$ = 5 events, one can observe the 2He + H and He + 3H channels, which saturate about 70\% for \emph{N}$_{tf}$ and about 50\% for \emph{N}$_{ws}$. Among other things, this circumstance may also reflect the effect of crossing the proton drip line in the fragmentation $^8$B $\rightarrow^7$B in neutron-stripping reactions [accompanied (or not accompanied) by the production of target fragments and by $^7$B decay to $^4$He + 3\emph{p} and 2$^3$He + \emph{p} states]. In neutron-stripping reactions, the energy threshold does not play a significant role, and the formation of these states may occur with commensurate intensities.

The statistics of 87 \emph{N}$_{tf}$ events characterized by \emph{Z}$_{pr}$ = 5 and $\Sigma$\emph{Z}$_{fr}$ = 4 (Table \ref{tab:1}) makes it possible to estimate the relative fractions of the proton-stripping reaction $^8$B $\rightarrow^7$Be and channels of higher multiplicity. This reaction was studied in measuring the momentum spectra of relativistic fragments of the $^7$Be nucleus \cite{Smedberg99,Cordina-Gil03}. In all, eight such events were found in the present analysis. It is noteworthy that their fraction in \emph{Z}$_{pr}$ = 5 and $\Sigma$\emph{Z}$_{fr}$ = 4 statistics is insignificant. Further, He + 2H events form approximately half the statistics of this class, and 2He and 4H follow them. The situation around the statistics of \emph{Z}$_{pr}$ = 5 and $\Sigma$\emph{Z}$_{fr}$ = 3 is similar. Here, there is no $^8$B $\rightarrow^6$Li event among 16 observed events. Thus, fragmentation involving proton stripping leads, as a rule, to the formation of He and H clusters rather than to the formation of single heavier nuclei.

Table \ref{tab:2} shows the distribution of the number of peripheral events, \emph{N}$_{pf}$, with $\Sigma$\emph{Z}$_{fr}$ = 5 for $^8$B nuclei over the numbers \emph{n}$_g$ and \emph{n}$_b$ of accompanying target-nucleus fragments. Under the conditions \emph{n}$_g$ = 1 and \emph{n}$_b$ = 0, the fraction of $^8$B $\rightarrow^7$Be + \emph{p} events associated with interactions on hydrogen nuclei is quite small. The observed distribution of the statistical sample over channels of fragment multiplicity \emph{n}$_b$ highlights the significance of \emph{N}$_{ws}$ events against \emph{N}$_{tf}$ events involving the overlap of the densities of colliding nuclei.

\begin{table}
\caption{\label{tab:2} Distribution of $^8$B-dissociation events over the
charge configurations at $\Sigma$\emph{Z}$_{fr}$ = 5 for various accompanying target-nucleus fragments}

\begin{tabular}{c|c|c|c|c|c|c|c}
\hline\noalign{\smallskip}

	~~\emph{n}$_g$~~  & ~~0~~ & ~~1~~ & ~~0~~ & ~~0~~ & ~~0~~ & ~~0~~ & ~~0~~  \\
    ~~\emph{n}$_b$~~  & ~~0~~ & ~~0~~ & ~~1~~ & ~~2~~ & ~~3~~ & ~~4~~ & ~~5~~  \\
\hline\noalign{\smallskip}
	~~He + 3H~~  &  ~~12~~  &  ~~6~~ & ~~8~~ & ~~3~~ & ~~2~~ & ~~3~~ & ~~-~~\\
	~~2He + H~~  &  ~~14~~  &  ~~3~~ & ~~8~~ & ~~2~~ & ~~4~~ & ~~-~~ & ~~1~~\\
	~~Be + H~~   &  ~~25~~  &  ~~1~~ & ~~3~~ & ~~3~~ & ~~1~~ & ~~-~~ & ~~-~~\\
	~~B~~      &  ~~1~~   &  ~~1~~ & ~~8~~ & ~~1~~ & ~~-~~ & ~~1~~ & ~~-~~\\

\hline\noalign{\smallskip}

\end{tabular}
%\hspace*{10cm}  % with the correct table height
\end{table}

\section{\label{sec:level3}ISOTOPIC COMPOSITION OF He AND H
RELATIVISTIC FRAGMENTS}

Of course, it would be of interest to identify completely the H and He fragments of the $^8$B nucleus via measuring their momenta by the multiple-scattering method. In employing the track-emulsion procedure, the fragment momenta \emph{p$\beta$c} can be determined by measuring the multiple scattering of fragments, but this is a difficult challenge. Relativistic hydrogen and helium isotopes are separated under the assumption that projectile fragments conserve the primary momentum per nucleon; that is, the fragment mass number is \emph{A}$_{fr}\approx$\emph{p$_{fr}\beta$c}/(\emph{p$_0\beta_0$c}). The respective event can then be interpreted in minute detail. Because of technical problems associated with the trackemulsion layers used and because of limitations associated with the angular scatter of fragment tracks,such measurements could not be performed for whole statistics.

\begin{figure}
\includegraphics[width=70mm]{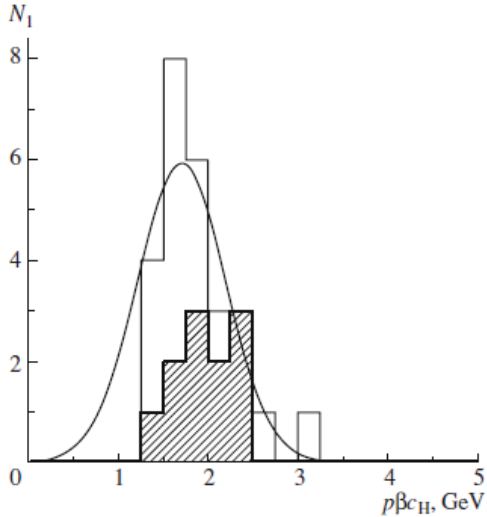}
\caption{\label{fig:3} Distribution of singly charged fragments of the $^8$B nucleus over the measured values of \emph{p$\beta$c} in the (15 tracks in white stars; unshaded histogram) Be + H and (11 tracks, including five tracks in white stars; shaded histogram) 2He + H dissociation channels. The solid curve represents an approximation by a Gaussian function peaked at 1.7 GeV.}
\end{figure}

For 26 singly charged fragments from $^8$B$\rightarrow$ Be + H and 2He + H events, the measured values of \emph{p$\beta$c}$_H$ are given in Fig. \ref{fig:3}. In those events, an H fragment is identified as a proton. The distribution is described by a single Gaussian function with a mean value of $\langle$\emph{p$\beta$c}$\rangle_H$ = 1.7 $\pm$ 0.2 GeV at $\sigma$ = 0.5 GeV. This corresponds to expected values.

\begin{figure}
\includegraphics[width=90mm]{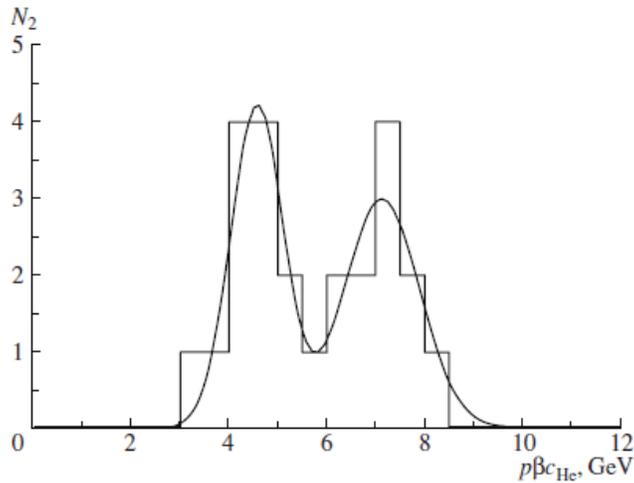}
\caption{\label{fig:4} Distribution of doubly charged fragments of the $^8$B nucleus over the measured values of \emph{p$\beta$c}$_{He}$. The solid curve represents an approximation by the sum of two Gaussian functions.}
\end{figure}

Figure \ref{fig:4} shows the distribution of the measured values of \emph{p$\beta$c}$_{He}$ for 24 doubly charged fragments chosen at random (twenty-two tracks from 2He + H events and two tracks from He + 3H events), which is satisfactorily described by the sum of of two normal distributions. The parameters of approximating functions correspond to values expected for the relativistic isotopes $^3$He and $^4$He, $\langle$\emph{p$\beta$c}$\rangle_{^3He}$ = 4.6 $\pm$ 0.2 GeV at $\sigma$ = 0.6 GeV and $\langle$\emph{p$\beta$c}$\rangle_{^4He}$ = 7.1 $\pm$ 0.3 GeV at $\sigma$ = 0.9 GeV. The distribution in question is indicative of an approximately identical fractions of $^3$He and $^4$He nuclei, as might have been expected for the $^8$B nucleus. The average values $\langle$\emph{p$\beta$c}$\rangle$ for $^1$H, $^3$He, and $^4$He are approximately in the same ratio as the fragment mass numbers \emph{A}$_{fr}$, and the values of $\sigma$ make it possible to separate the isotopes in question.

In four 2He + H events, \emph{p$\beta$c} could be measured for all tracks. Of these, three events are identified as $^4$He + $^3$He + $^1$H (these are events of the white-star type, events featuring \emph{n}$_b$ = 1 and \emph{n}$_g$ = 1 fragments, as well as \emph{n}$_b$ = 5 events), while one event was identified as $^3$He + $^3$He + $^1$H (event featuring an \emph{n}$_g$ = 1 fragment). Obviously, all of them comply with the assumption that it is precisely the $^8$B nucleus that undergoes dissociation.

Thus, the above results of measurement of the momenta \emph{p$\beta$c} confirm the identification of the primary isotope as a $^8$B nucleus and also demonstrate the potential of the method for a complete identification of the system of H and He fragments detected under conditions accepted for measurements.

\section{\label{sec:level4}ANALYSIS OF ANGULAR DISTRIBUTIONS}

\begin{figure}
\includegraphics[width=120mm]{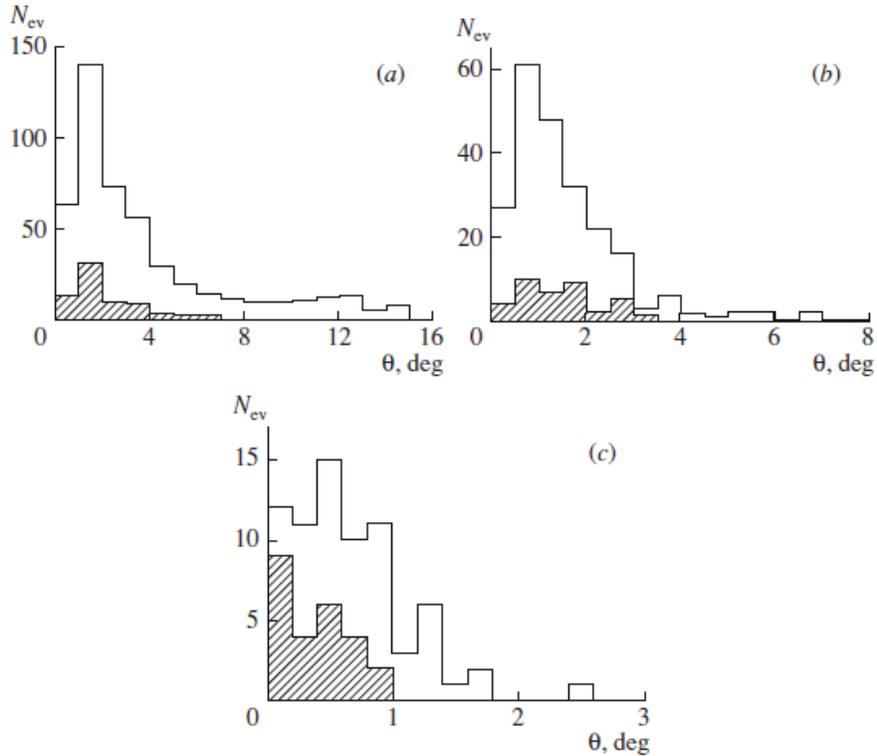}
\caption{\label{fig:5} Distributions of the polar emission angle è for relativistic fragments of charge (\emph{a}) \emph{Z}$_{fr}$ = 1 (479 tracks), (\emph{b}) \emph{Z}$_{fr}$ = 2 (224 tracks), and (\emph{c}) \emph{Z}$_{fr} > $2 (72 tracks). The shaded part of the histogram represents the contribution of white stars.}
\end{figure}

For \emph{N}$_{pf}$ events, Fig. \ref{fig:5} shows the distributions of measured polar emission angles $\theta$ for relativistic fragments of charge \emph{Z}$_{fr}$ = 1, 2, and \emph{Z}$_{fr} > $ 2. For \emph{Z}$_{fr}$ = 1 particles (Fig. 5a), the emission angles were measured over the range $\theta \leq$ 15$^{\circ}$. For the \emph{Z}$_{fr}$ = 1 fragments subjected to the analysis here, the maximum angle was chosen to be 8$^{\circ}$ in order to eliminate the contribution of participant protons. The angles for the \emph{Z}$_{fr}$ = 2 and \emph{Z}$_{fr} > $ 2 fragments were constrained by the condition $\theta\leq$ 6$^{\circ}$ (Fig. 5b) and the condition $\theta\leq$ 3$^{\circ}$ (Fig. 5c), respectively. Owing to the absence of limitations on the acceptance, these measurements may be of use in planning future experiments.

\begin{figure}
\includegraphics[width=90mm]{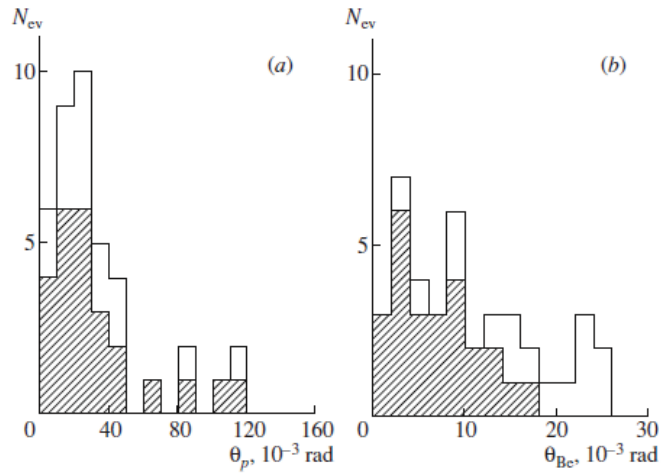}
\caption{\label{fig:6} Distribution of the polar emission angle $\theta$ in the reaction $^8$B$\rightarrow^7$Be + \emph{p} (40 events) for target-nucleus fragments of charge (\emph{a}) \emph{Z}$_{fr}$ = 1 and (\emph{b}) \emph{Z}$_{fr}$ = 4. The shaded part of the histogram represents the contribution of white stars (25 events).}
\end{figure}

For the $^7$Be + \emph{p} channel, Fig. \ref{fig:6} shows the distributions of the emission angle $\theta$ for projectile fragments of charge \emph{Z}$_{fr}$ = 1 (Fig. 6a) and \emph{Z}$_{fr}$  = 4 (Fig. 6b). These distributions are concentrated primarily in the region of small values. The mean value $\langle\theta\rangle$ for \emph{Z}$_{fr}$  = 1 fragments in white-star events is $\langle\theta_p\rangle$ = (33 $\pm$ 6) $\times$ 10$^{-3}$ rad, the root-mean-square deviation being RMS = 29 $\times$ 10$^{-3}$ rad. For \emph{Z}$_{fr}$  = 4 fragments, the mean value is $\langle\theta_{Be}\rangle$ = (6.9 $\pm$ 1.0) $\times$ 10$^{-3}$ rad and RMS = 4.5 $\times$ 10$^{-3}$ rad. The difference in the values of $\langle\theta\rangle$ reflects the difference in the fragment mass.

Measurements of the angle $\theta$ make it possible to calculate the transverse momenta \emph{P}$_T$ of relativistic fragments with a mass number \emph{A}$_{fr}$ according to the approximate relation \emph{P}$_T\approx $\emph{A}$_{fr}$\emph{P}$_0\sin\theta$. Going over to the reference frame comoving with the center of mass of the $^7$Be + \emph{p} system, one can compensate for the transverse-momentum transfer to the relativistic nucleus, and this leads to a \emph{P}$_T$ distribution characterized by the mean value of $\langle$\emph{P}$_T \rangle$ = 62 $\pm$ 11 MeV/\emph{c} (RMS = 54 MeV/\emph{c}). Thus, manifestations of a weak proton coupling to the core nucleus can be studied most comprehensively by analyzing the \emph{P}$_T$ distribution.

Table \ref{tab:3} presents the mean values of the aforementioned transverse momenta for events characterized by different topologies of target-nucleus fragmentation. An increase in the value $\langle$\emph{P}$_T(^8B^*)\rangle$ in events involving the excitation of the target nucleus in relation to the case of white stars can be noticed even within a small statistical sample. This increase is due to the growth of the mean momentum $\langle$\emph{P}$_{TBe}\rangle$ of the $^7$Be nucleus at an approximately constant mean momentum of the proton, $\langle$\emph{P}$_{Tp}\rangle$. This result suggests that the $^7$Be core undergoes scattering accompanied by the production of target-nucleus fragments and that there occurs an independent separation of a loosely bound proton. One can assume a two-step process of $^8$B dissociation followed by scattering. A sizable effect from the $^7$Be nucleus is determined by its large size. In our opinion, this observation is of use for obtaining deeper insight into reaction mechanism.

\begin{table}
\caption{\label{tab:3} Mean values (in MeV/\emph{c}) of the fragment transverse momenta, $\langle$\emph{P}$_T \rangle$, and and their vector sum,
$\langle$\emph{P}$_T(^8B^*) \rangle$, in the reaction $^8$B $\rightarrow^7$Be + \emph{p} for various sets of accompanying target fragments and in the absence of charged-meson production (\emph{n}$_s$ = 0).}

\begin{tabular}{c|c|c|c|c|c}
\hline\noalign{\smallskip}

	~~\emph{n}$_g$~~  & ~~0~~ & ~~1~~ & ~~0~~ & ~~0~~ & ~~0~~ \\
    ~~\emph{n}$_b$~~  & ~~0~~ & ~~0~~ & ~~1~~ & ~~2~~ & ~~3~~ \\
\hline\noalign{\smallskip}
    ~~Number of events~~&  ~~ ~~  &  ~~ ~~ & ~~ ~~ & ~~ ~~ & ~~ ~~\\
	~~$^7$Be + \emph{p}~~&  ~~25~~  &  ~~1~~ & ~~3~~ & ~~3~~ & ~~1~~ \\
	~~$\langle$\emph{P}$_{Tp}\rangle$~~ &  ~~66 $\pm$ 12~~   &  ~~38~~ & ~~52 $\pm$ 11~~ & ~~64 $\pm$ 12~~ & ~~65~~ \\
~~$\langle$\emph{P}$_{TBe} \rangle$~~      &  ~~97 $\pm$ 13~~   &  ~~130~~ & ~~301 $\pm$ 36~~ & ~~129 $\pm$ 60~~ & ~~298~~ \\
~~$\langle$\emph{P}$_T(^8B^*) \rangle$~~      &  ~~95 $\pm$ 15~~   &  ~~152~~ & ~~324 $\pm$ 45~~ & ~~170 $\pm$ 78~~ & ~~350~~ \\
\hline\noalign{\smallskip}

\end{tabular}
%\hspace*{10cm}  % with the correct table height
\end{table}

\begin{figure}
\includegraphics[width=90mm]{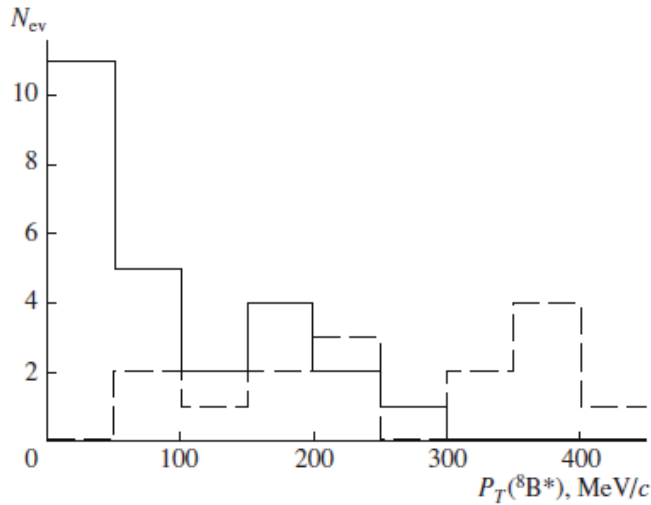}
\caption{\label{fig:7} Distribution of the total transverse momentum \emph{P}$_T(^8B^*)$ of $^7$Be + \emph{p} pairs produced in the reaction $^8$B$\rightarrow^7$Be + \emph{p}: (solid-line histogram) white stars (25 events) and (dashed-line histogram) interactions accompanied by target fragments and product particles (\emph{n}$_s$ + \emph{n}$_g$ + \emph{n}$_b$ $\neq$ 0; 15 events).}
\end{figure}

The distribution of the vector sum \emph{P}$_T(^8B^*$) of the transverse momenta of fragments originating from the $^7$Be + \emph{p} system may reflect the mechanisms of its production. Figure \ref{fig:7} shows the \emph{P}$_T(^8B^*$) distributions for events accompanied by target-nucleus fragments and for events not accompanied by them. The distribution corresponding to white stars has a mean value of $\langle \emph{P}_T(^8B^*) \rangle$ = 95$\pm$15 MeV/\emph{c} at RMS = 73 MeV/\emph{c}; for events featuring target-nucleus fragments and product mesons, $\langle \emph{P}_T(^8B^*) \rangle$ = 251 $\pm$ 29 MeV/\emph{c} at RMS = 112 MeV/\emph{c}. From a comparison of these distributions, one can draw the conclusion that the condition \emph{P}$_T(^8B^*) <$ 150 MeV/\emph{c} makes it possible to single out quite efficiently the kinematical region characteristic of the production of $^7$Be + \emph{p} white stars. The appearance of target fragments in events leads to a considerable increase in the corresponding values of $\langle \emph{P}_T(^8B^*) \rangle$.

\begin{figure}
\includegraphics[width=80mm]{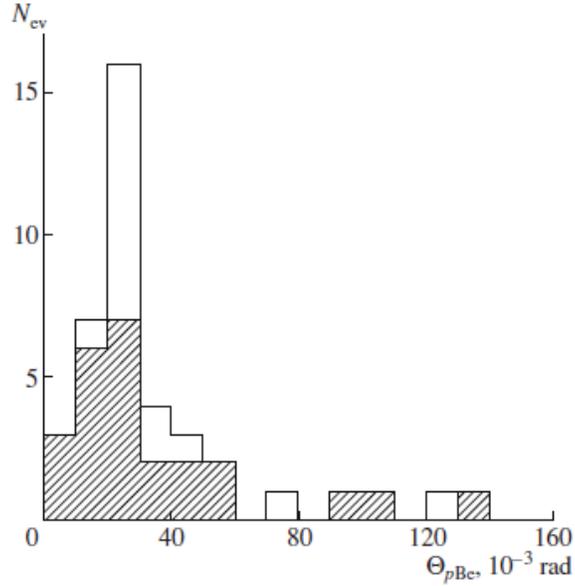}
\caption{\label{fig:8} Distribution of the $\Theta_{pBe}$ between the fragment momenta in the $^7$Be + \emph{p} channel (40 events). The shaded part of the histogram represents the contribution of white stars (25 events).}
\end{figure}

Figure \ref{fig:8} shows the distributions of the angles $\Theta_{pBe}$ between the fragment momenta in the $^7$Be + \emph{p} channel. For white stars, the mean value of the distribution is $\langle\Theta_{pBe}\rangle$ = (36 $\pm$ 6) $\times$ 10$^{-3}$ rad (RMS = 31 $\times$ 10$^{-3}$ rad). These distributions make it possible to estimate the invariant mass \emph{M}$^*$ of a fragment pair under the assumption that the primary momenta per nucleon \emph{P}$_0$ are conserved.

In order to estimate the excitation energy of the $^8$B nucleus, the distribution of the difference of the invariant mass \emph{M}$^*$ of the $^7$Be +\emph{p} system and the sum of the masses of the $^7$Be nucleus and proton, \emph{M}, is displayed in Fig. 9a. The mean value of this distribution is $\langle$\emph{Q}$_{pBe}\rangle$ = 4.3 ± 1.5 MeV (RMS = 7.6 MeV).The majority of the events are grouped in the range \emph{Q}$_{pBe} < $2 MeV. A more detailed pattern of the distribution in this range is
given in Fig. 9b. It includes 64\% of white stars (or 16 events) characterized by a mean value of $\langle$\emph{Q}$_{pBe}\rangle$ = 0.9 ± 0.1 MeV (RMS = 0.5 MeV). This value is compatible with the assumption of the decay of the $^8$B nucleus through the first excited state above the threshold for the $^7$Be + \emph{p} decay channel at 0.8 MeV with a width of 35 keV. The energy of the next excited state is 2.3 MeV.

\begin{figure}
\includegraphics[width=100mm]{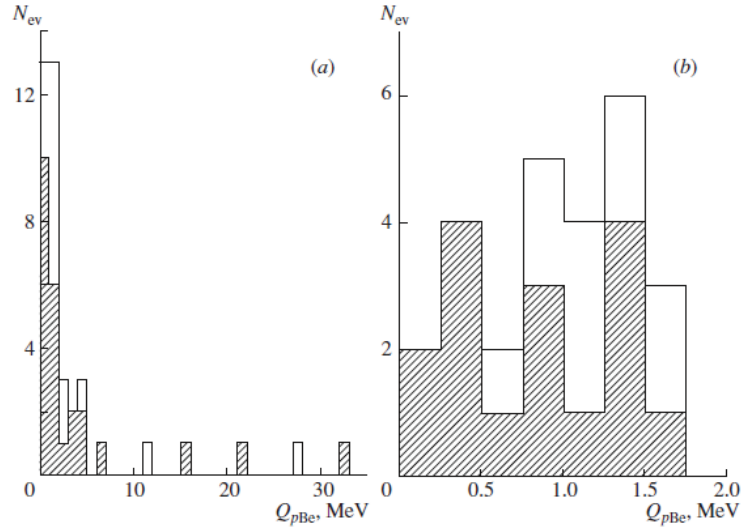}
\caption{\label{fig:9} Excitation-energy distributions for the channel $^8$B$\rightarrow^7$Be+\emph{p} in the ranges (\emph{a}) 0$-$35 MeV (40 events) and (\emph{b}) 0$-$2 MeV (25 events). }
\end{figure}

\section{\label{sec:level5}ESTIMATING THE CROSS SECTION FOR ELECTROMAGNETIC DISSOCIATION}

Products of the breakup process $^8$B$\rightarrow^7$Be + \emph{p} induced by a target-nucleus virtual photon should take the smallest values of the total transverse momentum. In view of this, a correlation in the azimuthal angle
$\varepsilon_{pBe}$ between these products should manifest itself in such events, as was previously indicated in \cite{Stanoeva07}. In selecting electromagnetic events, the soft criterion $\varepsilon_{pBe} > \pi$/2 may therefore be adopted in addition.

Figure 10a shows the distribution of the total transverse momentum for $^7$Be + \emph{p} white stars that satisfy the conditions \emph{Q}$_{pBe} < $ 5 MeV and $\varepsilon_{pBe} > \pi$/2. It contains 14 events characterized by a mean value of $\langle \emph{P}_T(^8B^*) \rangle$ = 50 $\pm$ 8 MeV/\emph{c} at RMS = 30 MeV/\emph{c}. The narrow angular distribution in Fig. 10b corresponds to these events, the parameters of this distribution being $\langle \theta_{^8B^*} \rangle$ = (3.1 $\pm$ 0.5) $\times$ 10$^{-3}$ rad and RMS = 1.9 $\times$ 10$^{-3}$ rad. It is precisely these events that correspond to the presumed criteria for the electromagnetic character of their formation and underlie the estimation of the cross section for electromagnetic dissociation through the $^7$Be + \emph{p} channel.

\begin{figure}
\includegraphics[width=100mm]{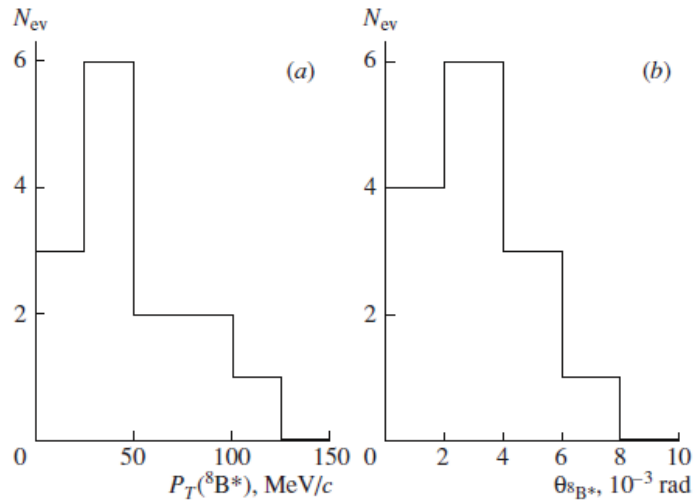}
\caption{\label{fig:10} Distribution of $^7$Be + \emph{p} white stars (14 events) in (\emph{a}) the total transverse momentum \emph{P}$_T(^8B^*$) and (\emph{b}) the angle $\theta_{^8B^*}$. The selections used are \emph{Q}$_{pBe} <$ 5 MeV and $\varepsilon_{pBe} > \pi$/2.}
\end{figure}

Because of a very strong dependence of the electromagnetic cross section on the target-nucleus charge in the form \emph{Z}$^2$, it is natural to assume a proportional contribution of Ag and Br nuclei from the composition of the nuclear track emulsion and to disregard the contribution of light nuclei. One can then associate nine $^7$Be + \emph{p} events with dissociation on Ag nuclei and approximately five events with dissociation on Br nuclei. Under this assumption, the cross section for their production can be estimated by the formula
\begin{center}
    \emph{$\sigma_Z$ = N$_{ev}$/(N$_Z$L)},
\end{center}
where \emph{N}$_{ev}$ is the number of events in this channel, \emph{N}$_Z$ is the concentration of atoms belonging to the type being considered (in cm$^{-3}$ units), and \emph{L} is the total number of the tracks within which \emph{N}$_{ev}$ events were found. The density of Ag atoms, as well as the density of Br atoms, is \emph{N}$_Z$ = 1.03 $\times$ 10$^{22}$ cm$^{-3}$. The scanned track length corresponding to the contribution of $^8$B nuclei in the composition of the beam is \emph{L} = 109.5 m, which leads to values of $\sigma_{Ag}$ = 81 $\pm$ 21 mb and $\sigma_{Br}$ = 44 $\pm$ 12 mb.

For electromagnetic dissociation on a Pb nucleus, an extrapolation of $\sigma_{Ag}$ leads to a value of $\sigma_{Pb}$ = 230 $\pm$ 60 mb, which is close to a theoretical value of about 210 mb \cite{Esbensen2000} and which is consistent with the assumption of the electromagnetic nature of selected
events. We can state that our approach to seeking and selecting $^7$Be + \emph{p} events of electromagnetic nature is quite efficient. Eleven $^7$Be + \emph{p} white stars that did not pass selections can be associated with the diffractive-dissociation contribution, which was discussed in \cite{Esbensen2000}. In order to estimate the cross sections in question to a higher precision, it is necessary to perform additional measurements that would enlarge statistics substantially.

For the $^4$He + $^3$He + \emph{p} channel, the number of events that satisfy the adopted criteria of electromagnetic dissociation and, hence, the partial cross section are approximately one-third as large as their counterparts for the $^7$Be + \emph{p} channel. We can only be confident of the fact of their generation. They can serve as a guideline for planning future experiments
involving a faster search for 2He + H events over layers of nuclear track emulsion—in particular, an identification of 2$^3$He + $^2$H events, which are the most interesting of them.

\section{\label{sec:level6}EVENTS OF $^{10}$C DISSOCIATION}
The formation of some $\Sigma$\emph{Z}$_{fr} >$ 5 white stars (Table \ref{tab:1}) may be due to the presence of an admixture of $^{10}$C nuclei in the composition of the beam used. Nuclei of the isotope $^{10}$C could be produced via the charge-exchange process $^{10}B \rightarrow ^{10}$C in the target intended for $^8$B production and then be captured into
the secondary beam because of a small difference in magnetic rigidity
from $^8$B (about 4\%) and the scatter of nuclei in momentum. The identification of primary track charges made it possible to separate \emph{Z}$_{pr}$ = 6 events. Their charge topology is spelled out in Table \ref{tab:4}. White stars characterized by $\Sigma$\emph{Z}$_{fr}$ = 6 do not contain \emph{Z}$_{fr} >$ 2 fragments. Their topology corresponds to the dissociation of the $^{10}$C  nucleus, which has a core in the form of $^8$Be, through the most probable channel $^{10}C \rightarrow ^8$Be + 2\emph{p}. A stronger disintegration of the carbon isotope is observed in events featuring target nucleus fragments.

\begin{table}
\caption{\label{tab:4} Distribution of the number (\emph{N}$_{pf}$) of peripheral interactions for \emph{Z}$_{pr}$ = 6 that were observed in a nuclear track emulsion exposed to a secondary beam of $^8$B nuclei over charge configurations}

\begin{tabular}{l|ccccc|cc}
\hline\noalign{\smallskip}
~~$\Sigma$\emph{Z}$_{fr}$~~&  &  & ~~\emph{N}$_z$~~ & & & ~~\emph{N}$_{ws}$~~ & ~~\emph{N}$_{tf}$~~\\
 & ~~5~~ & ~~4~~ & ~~3~~ & ~~2~~ & ~~1~~ & \\
\noalign{\smallskip}\hline\noalign{\smallskip}
~~7~~ & ~~-~~ & ~~-~~ & ~~-~~ & ~~1~~ & ~~5~~ & ~~-~~ & ~~1~~ \\
~~7~~ & ~~-~~ & ~~-~~ & ~~-~~ & ~~2~~ & ~~3~~ & ~~1~~ & ~~-~~ \\
~~6~~ & ~~-~~ & ~~-~~ & ~~-~~ & ~~2~~ & ~~2~~ & ~~3~~ & ~~5~~\\
~~6~~ & ~~-~~ & ~~-~~ & ~~-~~ & ~~1~~ & ~~4~~ & ~~-~~ & ~~10~~ \\
~~6~~ & ~~-~~ & ~~-~~ & ~~-~~ & ~~-~~ & ~~6~~ & ~~-~~ & ~~2~~ \\
~~5~~ & ~~-~~ & ~~-~~ & ~~-~~ & ~~2~~ & ~~1~~ & ~~-~~ & ~~5~~ \\
\noalign{\smallskip}\hline

\end{tabular}
%\hspace*{10cm}  % with the correct table height
\end{table}

Since these are the first observations of $^{10}$C-dissociation events as comprehensive as they are, we will present the features of one of such events. The angle between the tracks of He nuclei appearing as fragments is $\Theta_{2He}$ = 0.36$^{\circ}$, which, under the assumption that they are identical to $^4$He, leads to a difference of the effective mass and the doubled $^4$He mass of \emph{Q}$_{2He}$ = 0.17 MeV. This value satisfies the conditions of the breakup of the unbound nucleus $^8$Be. Near a He pair, there is a singly charged track corresponding to a proton. The pair angles between the tracks of the first alpha particle and the proton and between the tracks of the second alpha particle and the proton are $\Theta_{HeH}$  = 0.82$^{\circ}$ and $\Theta_{He2H}$  = 0.52$^{\circ}$, respectively. The difference of the effective mass and the sum of the rest masses of the 2$^4$He + 2\emph{p} system is then \emph{Q}$_{2HeH}$ = 0.44 MeV. Thus, this group of tracks may correspond to the decay $^9$B$ \rightarrow ^8$Be + \emph{p} of the unbound nucleus $^9$B formed upon the knockout of a proton from a $^{10}$C nucleus at a large angle, this being accompanied by the production of a pair of target-nucleus fragments. For the entire system, the excitation energy has a significant value of \emph{Q}$_{2HeH} \approx$ 13 MeV with \emph{P}$_T(^{10}$C$^*$) = 214 MeV/\emph{c}.

\section{\label{sec:level7}Conclusions}
The entire body of experimental data concerning the charges of beam-particle tracks; the charge topology of peripheral collisions; and angular features and identification of relativistic \emph{p}, $^3$He, and $^4$He fragments suggests the dominance of beam tracks of precisely $^8$B nuclei in the nuclear track emulsion used. Thus, it has been confirmed that the problem of irradiating nuclear track emulsions with $^8$B nuclei has been successively solved for the first time with the JINR nuclotron.

A systematic pattern of the distributions of charge combinations of fragments in the peripheral interactions of $^8$B nuclei in a nuclear track emulsion has been obtained. The main conclusion is that the contribution of the dissociation channel $^8$B $\rightarrow^7$Be + \emph{p} is dominant in events that do not involve the production of target-nucleus fragments or mesons.

An analysis of angular correlations in events involving target-nucleus fragments and in events not involving such fragments made it possible to
validate selections of events of the electromagnetic dissociation process $^8$B $\rightarrow^7$Be + \emph{p} in the total transverse momentum of fragments, \emph{P}$_T$($^8$B$^*$) $<$ 150 MeV/\emph{c}; the pair excitation energy, \emph{Q}$_{pBe} <$ 5 MeV; and the azimuthal angle of divergence of $^7$Be and \emph{p} fragments, $\varepsilon_{pBe} > \pi$/2. An estimate of the cross section for electromagnetic dissociation on nuclei from the composition of the track emulsion used has been obtained for the first time for $^7$Be + \emph{p} events that does not involve the production of target-nucleus fragments or product mesons. An extrapolation of this estimate to a lead nucleus suggests agreement with the results of theoretical calculations.

Events of the peripheral dissociation of the $^{10}$C nucleus have been observed for the first time. An indication that the dissociation channel $^{10}$C $\rightarrow$ 2He + 2H is dominant in events that do not involve the production of target-nucleus fragments or charged mesons has been obtained. The identification of the isotope $^{10}$C gives grounds to hope that a $^{10}$C beam can be formed in the charge-exchange reaction $^{10}$B $\rightarrow$ $^{10}$C under conditions convenient for studies in track-emulsion experiments.

The conclusions drawn in this study are of value for a comparative analysis of results obtained recently in exposures of nuclear track emulsions to $^9$C and $^{12}$N nuclei in secondary beams from the JINR nuclotron.

\begin{acknowledgments}
We are grateful to I.I. Sosul'nikova, À.M. Sosul'nikova,and G.V. Stel'makh from JINR for painstaking effort in data accumulation.

This work was supported by the Russian Foundation for Basic Research (project nos. 96-1596423, 02-02-164-12a, 03-02-16134, 03-02-17079, 04-02-17151, and 04-02-16593), VEGA grants (nos. 1/2007/05 and 1/0080/08) from the Agency for
Science of the Ministry for Education of the Slovak Republic and the Slovak Academy of Sciences, and grants from the JINR Plenipotentiaries of the
Republic of Bulgaria, the Slovak Republic, the Czech Republic, and Romania in 2002$-$2008.
\end{acknowledgments}


\begin{thebibliography}{}
\bibitem{Aumann05}
T. Aumann, Eur. Phys. J. \textbf {A26}, 441 (2005).
\bibitem{Bertulani88}
C. A. Bertulani and G. Baur, Phys. Rep. \textbf{163}, 299 (1988).
\bibitem{Baur96}
G. Baur and H. Rebel, Annu. Rev. Nucl. Part. Sci. \textbf{46}, 321 (1996).
\bibitem{nuclth/9710060}
G. Baur, S. Typel, and H. H. Wolter, nuclth/9710060.
\bibitem{Baur03}
G. Baur, K. Hencken, and D. Trautmann, Prog. Part. Nucl. Phys. \textbf{51}, 487 (2003); nucl-th/0304041.
\bibitem{Iwasa99}
N. Iwasa et al., Phys. Rev. Lett. \textbf{83}, 2910 (1999).
\bibitem{web}
The BECQUEREL Project, http://becquerel.jinr.ru/
\bibitem{Adamovich99}
M. I. Adamovich et al., Yad. Fiz. \textbf{62}, 1461 (1999)
[Phys. At. Nucl. \textbf{62}, 1378 (1999)].
\bibitem{Andreeva05}
N. P. Andreeva et al., Yad. Fiz. \textbf{68}, 484 (2005) [Phys.
At. Nucl. \textbf{68}, 455 (2005)]; N. P. Andreeva et al., nuclex/
0605015.
\bibitem{Artemenkov08}
D. A. Artemenkov et al., Yad. Fiz. \textbf{71}, 1595 (2008)
[Phys. At. Nucl. \textbf{71}, 1565 (2008)].
\bibitem{Esbensen2000}
H. Esbensen and K. Hencken, Phys. Rev. C \textbf{61}, 054606 (2000).
\bibitem{Stanoeva07}
R. Stanoeva et al., Yad. Fiz. \textbf{70}, 1255 (2007) [Phys.
At. Nucl. \textbf{70}, 1216 (2007)]; R. Stanoeva et al., nuclex/
0605013.
\bibitem{Rukoyatkin06}
P. A. Rukoyatkin, L. N. Komolov, R. I. Kukushkina,
and V. N. Ramzhin, Czech. J. Phys. Suppl. C \textbf{56},
C379 (2006).
\bibitem{Smedberg99}
M. H. Smedberg et al., Phys. Lett. B \textbf{452}, 1 (1999).
\bibitem{Cordina-Gil03}
D. Cordina-Gil et al., Nucl. Phys. A \textbf{720}, 3 (2003).
\end{thebibliography}
\end{document}